\documentclass[twocolumn,showpacs,preprintnumbers,amsmath,amssymb]{revtex4}

\usepackage{graphicx}
\usepackage{dcolumn}
\usepackage{bm}

\begin{document}

\preprint{arXiv preprint}

\title{Resonance Fluorescence from a Coherently Driven Semiconductor Quantum Dot in a Cavity}

\author{Andreas Muller}
\author{Edward B. Flagg}%
\author{Pablo Bianucci}%
\author{Xiaoyong Wang}%
\author{Chih-Kang Shih}%
 \email{shih@physics.utexas.edu}
\affiliation{%
Department of Physics, The University of Texas at Austin, Austin, TX 78712 
}%

\author{Dennis G. Deppe}
\affiliation{
College of Optics and Photonics (CREOL), University of Central Florida, Orlando, FL 32816
}%

\author{Wenquan Ma}
\author{Jiayu Zhang}
\author{Min Xiao}
\author{Gregory J. Salamo}
\affiliation{
Department of Physics, University of Arkansas, Fayetteville, AR 72701
}%

\date{\today}

\begin{abstract}
We show that resonance fluorescence, i.e. the {\em resonant} emission of a {\em coherently} driven two-level system, can be realized with a semiconductor quantum dot. The dot is embedded in a planar optical micro-cavity and excited in a wave-guide mode so as to discriminate its emission from residual laser scattering. The transition from the weak to the strong excitation regime is characterized by the emergence of oscillations in the first-order correlation function of the fluorescence, $g(\tau)$, as measured by interferometry. The measurements correspond to a Mollow triplet with a Rabi splitting of up to 13.3 $\mu$eV. Second-order-correlation measurements further confirm non-classical light emission.
\end{abstract}

\maketitle

Semiconductor quantum dots (QDs) \cite{bayer2000hse} have offered unique opportunities to investigate sophisticated quantum optical effects in a solid-state system. These include quantum interference \cite{bonadeo1998coc}, Rabi oscillations \cite{htoon2002iro, kamada2001ero, patton2005cca, stievater2001roe, zrenner2002cpt}, as well as photon anti-bunching \cite{michler2000qds}, and were previously only observable in isolated atoms or ions. In addition, QDs can be readily integrated into optical micro-cavities making them attractive for a number of applications, particularly quantum information processing and high efficiency light sources. For example, QDs could be used to realize deterministic solid-state single photon sources \cite{darquie2005csp, mckeever2004dgs, keller2004cgs} and qubit-photon interfaces \cite{yao2005tcs}. Advances in high-Q cavities have shown that not only can the spontaneous emission rate be dramatically increased by the Purcell effect \cite{gerard1998ese, deppe1999ese}, but emission can be reversed in the strong coupling regime \cite{yoshie2004vrs, reithmaier2004scs, hennessy2007qns}. Despite these efforts, however, quantum dot-based cavity quantum electrodynamics (QED) lacks an ingredient essential to the success of atomic cavity QED, namely the ability to truly resonantly manipulate the two-level system \cite{darquie2005csp, mckeever2004dgs, keller2004cgs}. Current approaches can at best populate the dot in one of its excited states, which subsequently relaxes in some way to the emitting ground state. This incoherent relaxation has been addressed theoretically \cite{kiraz2004qds, fattal2006csp},  and experimentally \cite{santori2002ips} but direct resonant excitation and collection in the ground state has so far not been reported as it is very challenging to differentiate the resonance fluorescence from same-frequency laser scattering off defects, contaminants, etc. In quantum dots without cavities, coherent manipulation of ground-state excitons has nonetheless been achieved with a number of techniques including differential transmission \cite{stievater2001roe}, differential reflectivity \cite{unold2004ose}, four-wave mixing \cite{borri2001udt}, photodiode spectroscopy \cite{zrenner2002cpt}, and Stark-shift modulation absorption spectroscopy \cite{alen2003ssm}. However, none of these is able to collect and use the actual photon emission which limits their use in many potential applications of QDs.

This report presents the first measurement of resonance fluorescence in a single self-assembled quantum dot. Described by Mollow in 1969 \cite{mollow1969}, the resonant emission of a two-state quantum system under strong coherent excitation is distinguished by an oscillatory first-order correlation function, $g(\tau)$, that we observe with interferometry. We use a planar optical micro-cavity to guide the excitation laser between the cavity mirrors and simultaneously enhance the single photon emission in the orthogonal direction. Overcoming previous limitations associated with incoherent excitation, our approach enables, for the first time, true resonant excitation of a single dot in a cavity.

\begin{figure}
\includegraphics[width=3.in]{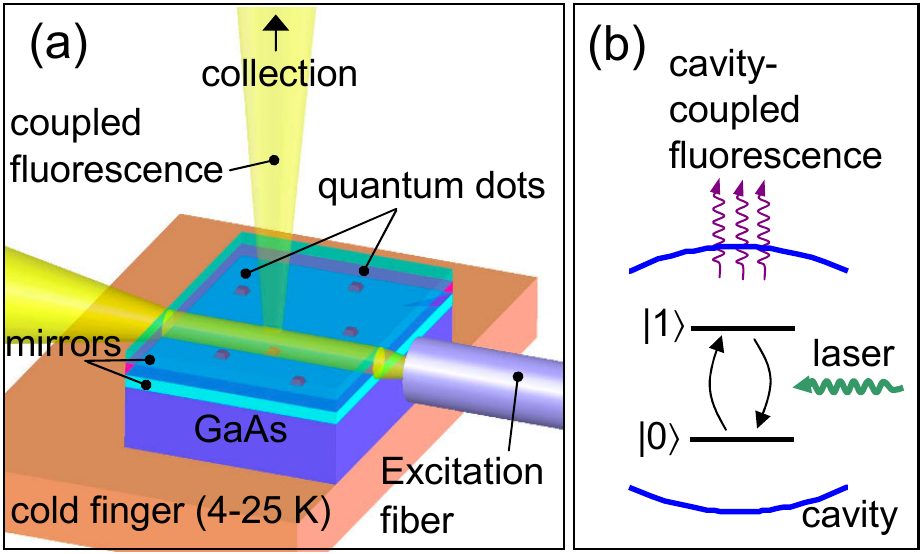}
\caption{\label{fig1:epsart} (color online). (a) Apparatus for orthogonal excitation and detection. (b) Energy level diagram for two-level quantum dot. The two arcs represent the micro-cavity in which the dots are embedded.}
\end{figure}

\begin{figure*}
\includegraphics[width=4.5in]{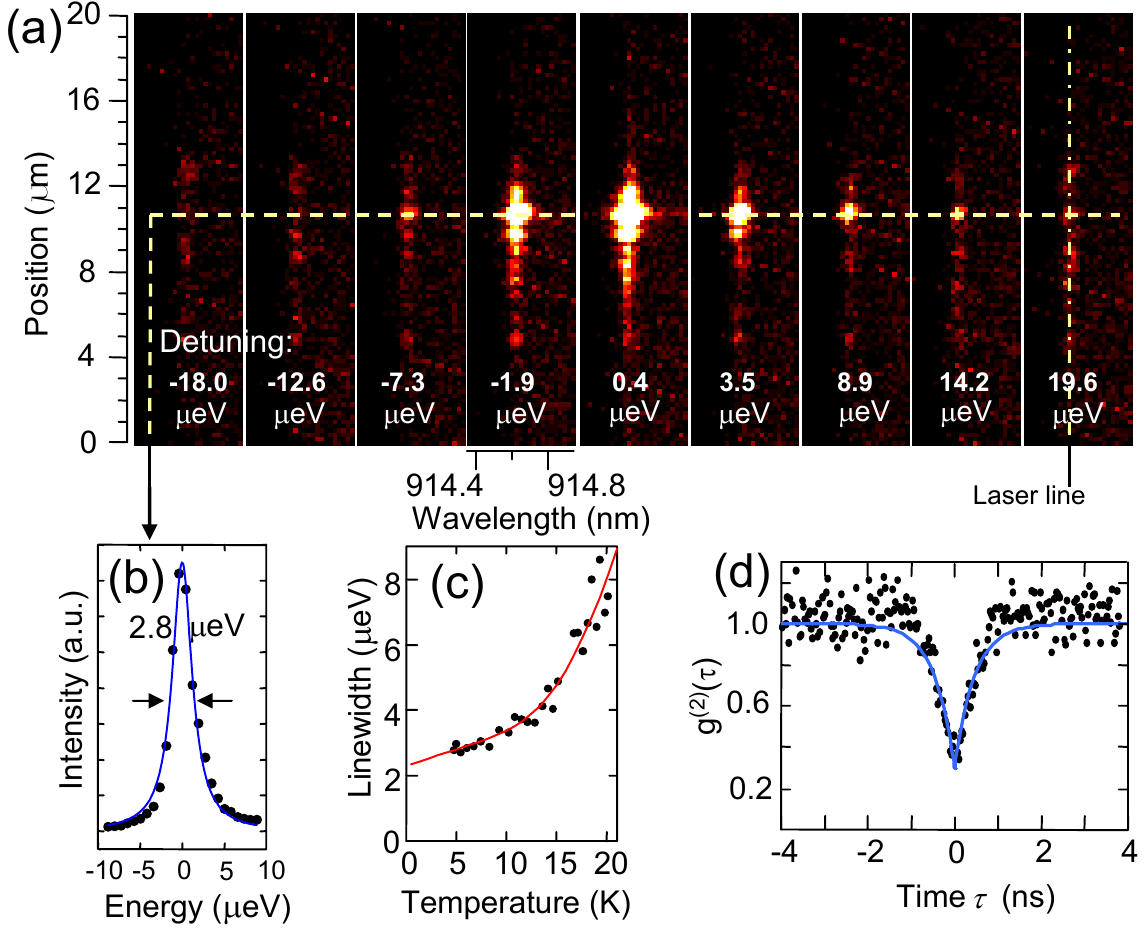}
\caption{\label{fig2:wide} (color online). (a) Spatially (ordinates) and spectrally (abscissas) resolved fluorescence images for a single resonantly excited quantum dot. The residual laser appears as a faint line. (b) Spectral line profile of dot in (a). (c) Linewidth temperature dependence. (d) Second-order correlation measurement of another single quantum dot in resonance fluorescence.}
\end{figure*}

Self-assembled InGaAs QDs were grown epitaxially between two distributed Bragg reflectors of moderate reflectivity (Fig.  \ref{fig1:epsart}). While the sample is maintained at low temperature in a He flow cryostat, a single mode optical fiber, mounted on a three-axis inertial walker at room temperature, is brought within a few microns of the cleaved sample edge. An in-plane polarized tunable continuous-wave Ti:Sapphire laser is introduced through the fiber to excite the dots; it couples efficiently into the high index semiconductor and propagates deeply before diverging appreciably. The QD emission is then collected by a conventional micro-PL setup equipped with a two-dimensional charge coupled device (CCD) detector mounted on an imaging spectrograph. We focus here on QDs coupled to a cavity mode centered around 915 nm, with a quality factor of about 250. For first-order correlation measurements, a Mach-Zehnder interferometer is inserted into the collection beam path. When the laser is frequency-scanned over the excitonic ground-state of a single QD, the resonance fluorescence is observed as a bright peak in the CCD images, localized both spectrally and spatially. In contrast, the remaining background laser light appears as a faint vertical (i.e. spatially delocalized) line. In Fig. \ref{fig2:wide}(a), a series of such CCD images at increasing excitation energy are shown. The laser bandwidth is less than 40 MHz, narrow enough that the total integrated intensity as a function of detuning measures the homogeneous linewidth of the ground state transition, as plotted explicitly in Fig.  \ref{fig2:wide}(b). For this particular dot we obtain a full width at half maximum (FWHM) of 2.8 $\mu$eV ($T_2$=470 ps) at 4.7 K. A strong dependence on temperature is observed [Fig. \ref{fig2:wide}(c)] and all subsequent measurements are performed at 10 K. Moreover, second-order correlation measurements, performed on single peaks using a Hanbury-Brown and Twiss setup (HBT) \cite{michler2000qds}, reveal a pronounced anti-bunching dip [Fig. \ref{fig2:wide}(d)], confirming their single emitter nature.

\begin{figure}
\includegraphics[width=2.6in]{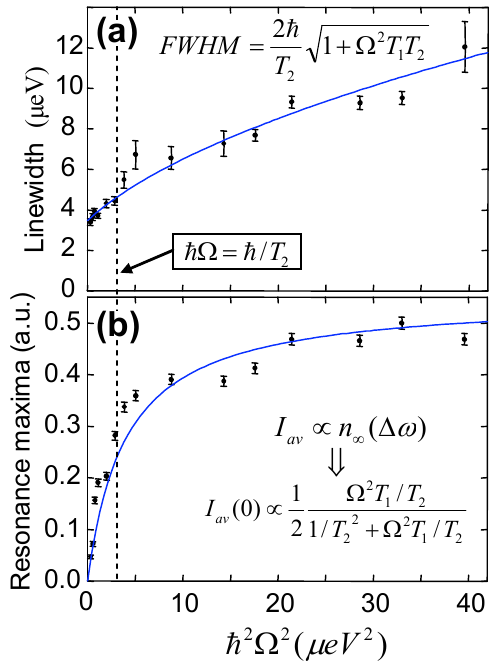}
\caption{\label{fig3:epsart} (color online). (a) Linewidth as a function of square Rabi energy. The resonances were recorded by scanning the laser over the quantum dot transition, and monitoring the total emitted intensity. (b) Resonance amplitude versus square Rabi energy extracted from the same data. The solid lines are obtained using the theoretically predicted formulas for the steady state population as a function of detuning following Eq. (2) and Eq. (4).}
\end{figure}

The interaction of a single QD with an external, near resonant electric field is described by the two-level optical Bloch equations in which the field is treated semi-classically and the dipole approximation is assumed \cite{scully}. Assessing the validity of this description, particularly under strong excitation, characterizes much of the progress in coherent QD spectroscopy in the past decade. Milestone experiments include the demonstration of quantum interference \cite{bonadeo1998coc}, Rabi oscillations \cite{stievater2001roe, zrenner2002cpt, kamada2001ero, htoon2002iro, patton2005cca, borri2002roe}, the optical Stark effect \cite{unold2004ose}, Ramsey Fringes \cite{stufler2006rfe}, as well as multilevel manipulation schemes such as two-photon Rabi oscillations \cite{li2003aoq}. At low intensity, a harmonic driving field (amplitude $E_0$), which may be detuned from the transition frequency of the two-level system by an amount $\Delta\omega=\omega-\omega_0$, initially increases the population of the upper state. When the field is so strong that the Rabi frequency $\Omega=\mu E_0$ exceeds the total decoherence rate $1/T_2$ in the system, however, the probability to find it in the upper state reaches a maximum before it decreases again. Here $\mu$ denotes the dipole moment of the transition with resonance frequency $\omega_0$. In fact, both the populations and the coherences of the system then oscillate at the Rabi frequency, which in the language of quantum computation corresponds to quantum bit rotations. Written out explicitly in the rotating wave approximation, the Bloch equations for the upper and lower state populations, $n(t)=\text{Tr}\{\rho(t)|1\rangle\langle1|\}$ and $m(t)=\text{Tr}\{\rho(t)|0\rangle\langle0|\}$, and for the coherence, $\alpha(t)=\text{Tr}\{\rho(t)|0\rangle\langle1|\}$, read:
\begin{equation}
\begin{array}{l}
\frac{d}{dt} n(t)=-i \frac{\Omega}{2}(\alpha(t)-\alpha^\ast(t))-\frac{n(t)}{T_1}  \\
\\
\frac{d}{dt} \alpha(t)=-i \frac{\Omega}{2}(n(t)-m(t))+i\alpha(t)\Delta\omega-\frac{\alpha(t)}{T_2}   \\
\end{array}
\label{molloweq1}
\end{equation}
Here $\rho(t)$ is the density operator, and $T_1$ and $T_2$ denote the diagonal and off-diagonal phenomenological damping constants, respectively, and $n(t)+m(t)=1$. The quasi steady-state solutions of Eqs. \ref{molloweq1} are obtained as:
\begin{equation}
\begin{array}{l}
n_\infty(\Delta\omega)=\frac{1}{2}\frac{\Omega^2 T_1/T_2}{\Delta\omega^2+T_2^{-2}+\Omega^2 T_1/T_2} \\
\\
\alpha_\infty(\Delta\omega)=\frac{i\Omega}{2} \frac{1/T_2+i\Delta\omega}{\Delta\omega^2+T_2^{-2}+\Omega^2 T_1/T_2} \\
\end{array}
\label{molloweq2}
\end{equation}
and describe well-known saturation phenomena which are directly observed in the experiments with single dots, since the time-averaged fluorescence intensity is proportional to $n_\infty(\Delta\omega)$. Specifically, one can see that (i) the total integrated fluorescence at resonance ($\Delta\omega=0$) saturates once the square Rabi frequency substantially exceeds the quantity $(T_1 T_2)^{-1}$, (ii) that the linewidth of the Lorentzian in Eq. (\ref{molloweq2}) increases slowly with the square root of intensity, a phenomenon known as Òpower broadeningÓ, and (iii) that the low intensity limit of the linewidth equals $2/T_2$.

More interesting is the actual shape of the fluorescence spectrum, which goes beyond the straightforward steady-state solutions. In fact, while the optical Bloch equations are directly borrowed from nuclear magnetic resonance theory, a comprehensive theoretical description of resonance fluorescence was only given in 1969 by Mollow \cite{mollow1969}. He first obtained the two-time (first-order) correlation function $g(t,\tau)=\langle b^\dag (t) b(t+\tau)\rangle$ of the field emitted by the system, where $b$ and $b^\dag$ are the field operators which are proportional to the atomic dipole operators |$|0\rangle\langle1|$ and $|1\rangle\langle0|$ \cite{mollow1969}. The complete resonance fluorescence spectrum was then derived as the Fourier transform of $g(t,\tau)$ and results in the well-known "Mollow triplet". Reduction to single time expectation values is then done with the quantum regression theorem \cite{scully}. Here we use a Mach-Zehnder interferometer to measure the correlation function directly, obtained as:
\begin{equation}
\begin{array}{l}
g(\tau)=|\alpha_\infty(0)|^2+\frac{n_\infty(0)}{2}e^{-\tau/T_2}+ \\
\\
+n_\infty(0) e^{-\tau(1/T_1+1/T_2)/2}\{N \cos(\Omega'\tau)+M\sin(\Omega'\tau)\}   \\
\end{array}
\label{molloweq3}
\end{equation}
where $N$ and $M$ denote constants that depend on $T_1$, $T_2$, and $\Omega$, and $\Omega'=\sqrt{\Omega^2-(1/T_1-1/T_2)^2/4}$. When $\Omega\ll1/T_2$, then $g(\tau)$ reduces to a simple exponential decay, with decay constant $T_2$, corresponding to a Lorentzian spectral line profile of FWHM $2/T_2$. On the other hand, when $\Omega\gg1/T_2$, the system is in the strong excitation regime and  $g(\tau)$ is oscillatory. Note that $T_1$ and $T_2$ are related through $T_2^{-1}=(2T_1)^{-1}+\gamma$, where $\gamma$ denotes pure dephasing (i.e. loss of coherence without population decay). 

For the same dot as in Fig. \ref{fig2:wide}(a), the excitation linewidth is plotted as a function of intensity in Fig. \ref{fig3:epsart}(a) to illustrate power broadening. As mentioned in the discussion of Eq. (\ref{molloweq2}), the value of $T_2$ can be obtained from the low intensity limit of the linewidth; here $T_2=380$ ps. If $T_1$ is also known, these measurements provide the proportionality constant between the excitation intensity and the square Rabi frequency. At zero temperature there would be no dephasing due to phonons, and the low intensity resonant pump should not cause spectral broadening by creating nearby transient charges. Therefore $T_1$ can be taken from the $T\rightarrow0$ limit of the linewidth in Fig. \ref{fig2:wide}(c), which means $T_1$=290 ps. With this extrapolated $T_1$ and measured $T_2$, we plot our data as a function of $\Omega^2$ directly. The emission intensity as a function of excitation intensity is plotted in Fig.  \ref{fig3:epsart}(b) and clearly shows the population saturation behavior predicted by Eq. (\ref{fig2:wide}) at intensities such that $\Omega\gg(T_1 T_2)^{-1/2}$. Noted on the graphs of Fig. \ref{fig3:epsart} are the thresholds for reaching the strong excitation regime. 

\begin{figure}
\includegraphics[width=2.7in]{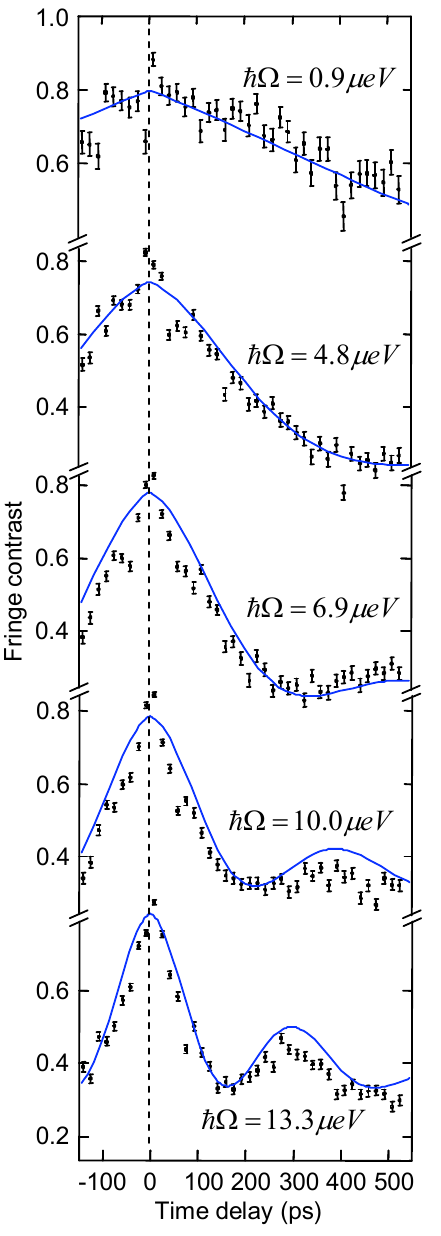}
\caption{\label{fig4:epsart} (color online). Fringe contrast as a function of time delay of the resonance fluorescence from a single quantum dot, for a range of excitation intensities. The corresponding Rabi energies were extracted from the oscillations, and extrapolated to weaker intensities. Fits represent the first order correlation function of Eq. (5).}
\end{figure}

To measure $g(\tau)$, the fluorescence is interfered with itself [Fig. \ref{fig4:epsart}(a)]. The length of one interferometer arm, and thus the corresponding time delay, is varied coarsely over a number of points with a highly stable translation stage, and finely using a piezo actuator to record the fringe contrast at each point [inset of Fig. \ref{fig4:epsart}(a)]. The contrast is defined as the difference between the maximum and minimum of the interference signal divided by their sum, and as a function of time delay corresponds to $g(\tau)$.  Such a technique, borrowed from Fourier spectroscopy \cite{santori2002ips}, can routinely provide an equivalent spectral resolution of about 1 $\mu$eV, much smaller than is available with conventional grating-based spectrometers. For the same dot as in Fig. \ref{fig2:wide} and Fig. \ref{fig3:epsart}, we examine the resulting fringe contrast as a function of time delay; it is plotted in Fig. \ref{fig4:epsart} for various excitation intensities, i.e. Rabi energies. Fits to the data are plotted on top of the data points using Eq. (5) and the above values of $T_1=$290 ps and $T_2=$380 ps. The Rabi frequency, $\Omega$, as well as an offset due to laser background are chosen to best fit the data; Rabi energies up to 13.3 $\mu$eV are found. 

When the intensity is weak, i.e. $\Omega\ll1/T_2$, a single exponential decay is obtained. But with increasing intensity, the fringe contrast develops an oscillatory feature at frequency $\Omega'$, defined above, which approximately equals $\Omega$ if $\Omega\gg1/T_2$. The oscillations can be understood as an amplitude modulation imposed on the field by the two-level system undergoing fast Rabi cycles. The system is sufficiently coherent to observe several oscillations, corresponding to a distinctive Mollow triplet in the frequency domain. The onset of these oscillations is qualitatively consistent with the steady state measurements of Fig. \ref{fig3:epsart}: once saturation has occurred, the two-level system is in the strong excitation regime. The simple two-level analysis with phenomenological $T_1$ and $T_2$ provides a satisfactory agreement considering the imitations of this model well-known in the literature, for example in the description of Rabi oscillations \cite{forstner2003pad, machnikowski2004rnp}.

In addition to their historical importance in early resonance fluorescence theory, continuous wave measurements, unlike pulsed measurements, also offer a distinct signature in the second-order correlation function under strong excitation. As reported in atomic experiments \cite{diedrich1987nrs} and described by an extension to MollowÕs calculation, the coherent oscillations profoundly modify the usual anti-bunching trace measured with an HBT setup. When sufficiently coherent, i.e. $\Omega\gg1/T_2$, it in fact resembles the correlation function that an ultra-fast pulsed emitter would provide. Future experiments with improved timing resolution will enable the measurement of these modulations with QDs. Nonetheless, with the exception of the completely non-classical anti-bunching obtained here at low-intensity [Fig. \ref{fig2:wide}(d)], the $g^{(1)}(\tau)$ measurements of Fig. \ref{fig4:epsart} provide no less information than those of $g^{(2)}(\tau)$. The first- and second-order correlations are complementary in that they illustrate a wave or particle picture of light, respectively. Finally, we anticipate that the capabilities demonstrated here open up avenues for probing fundamental phenomena in QDs such as squeezing. The latter involves interfering the fluorescence with a reference laser \cite{scully} and is not possible under incoherent excitation. Using a three dimensional microcavity instead of a simple planar structure would further enable advanced experiments relying on cavity quantum electrodynamic effects, many of which have been proposed for quantum information processing applications and realized in atomic systems. This might be achieved straightforwardly with all-epitaxial microcavities \cite{muller2006saa}, that possess a bulk morphology ideally suited to introduce a wave-guided laser.

In conclusion, our measurements, in which QDs are laterally excited in a microcavity, realize the goal of resonant coherent control of the excitonic ground state while simultaneously collecting its fluorescence. Using this method in concert with interferometry we achieve the first observation, in a driven solid-state two-level system, of resonance fluorescence in the strong excitation regime. Single photon emission is further confirmed by pronounced anti-bunching. Background-free resonant measurements offer new coherent control capabilities resembling those available for manipulating trapped atoms and ions, yet in a monolithic, scalable solid-state system.

\begin{acknowledgments}
The authors thank Hailin Wang, David Press, Wang Yao and Elaine Li for fruitful discussions, and acknowledge financial support from the National Science Foundation (DMR-0210383, DMR-0606485 and DGE-054917), and the W. Keck foundation.
\end{acknowledgments}

\bibliography{Mollow}

\end{document}